\documentclass[letters,usenatbib]{mnras}
\usepackage{newtxtext,newtxmath}

\usepackage[T1]{fontenc}
\usepackage{orcidlink}
\DeclareRobustCommand{\VAN}[3]{#2}
\let\VANthebibliography\thebibliography
\def\thebibliography{\DeclareRobustCommand{\VAN}[3]{##3}\VANthebibliography}

\usepackage{graphicx}	
\usepackage{amsmath}	

\title[Primordial gravitational waves in DHOST inflation]{Primordial gravitational waves in DHOST inflation}

\author[P. Brax and A. Lazanu]{
Philippe Brax$^{\orcidlink{0000-0003-0727-3186}{1}}$\thanks{E-mails: philippe.brax@ipht.fr (PB); andrei.lazanu@manchester.ac.uk (AL)}
and Andrei Lazanu$^{\orcidlink{0000-0002-8061-9828}{2}}$\footnotemark[1]
\\
$^{1}$Institut de Physique Th\'eorique, Universit\'e Paris-Saclay, CEA, CNRS, F-91191 Gif-sur-Yvette Cedex, France\\
$^{2}$Department of Physics and Astronomy, University of Manchester, Manchester M13 9PL, United Kingdom
}

\date{Accepted 2025 April 24. Received 2025 April 17; in original form 2025 March 05.}

\pubyear{\the\year{}}

\begin{document}
\label{firstpage}
\pagerange{\pageref{firstpage}--\pageref{lastpage}}
\maketitle

\begin{abstract}
We consider DHOST inflationary models with a shift symmetry leading to a de Sitter space-time at the background cosmological level. Deviations from scale invariance of the scalar and tensor perturbations follow from the breaking of the shift symmetry by quadratic and quartic operators.  These models show a strong violation of the consistency relation of single-field inflationary models with a very flat spectrum of tensor perturbations. This opens up the prospects of future detection of primordial gravitational waves by mHz experiments.
\end{abstract}

\begin{keywords}
gravitational waves -- early Universe -- inflation
\end{keywords}



\section{Introduction}

The $\Lambda$CDM model has long been the most successful description of  the current Universe, including the Cosmic Microwave Background (CMB) and the late-time acceleration. The model is based on an initial ``Big Bang'', followed by a short period of accelerated expansion and then by slower expanding radiation and matter epochs, and then again by an accelerated expansion period, dominated by dark energy. Both the inflationary and late-time accelerating epochs are currently unexplained and are suitable for parametrisations with scalar fields. In this context, scalar-tensor theories have been developed, out of which Degenerate Higher-Order Scalar-Tensor (DHOST) Theories are the most general theories involving a scalar field leading to the absence of  Ostrogradski ghost instabilities  \citep{Langlois:2015cwa, Achour:2016rkg, BenAchour:2016fzp, Crisostomi:2016czh, Crisostomi:2018bsp, Bombacigno:2021bpk}. They generalise the Horndeski \citep{Horndeski1974} and beyond-Horndeski \citep{Zumalacarregui:2013pma,Gleyzes:2014dya, Gleyzes:2014qga} theories, and they have been used to develop models describing both the late-time cosmic acceleration and inflation \citep{Brax:2021qlx,Brax:2021bok,Sohn:2023dbp} (see \citet{Langlois:2018dxi,Kobayashi:2019hrl,Lazanu:2024mzj} for reviews). The formalism for determining the power spectrum of primordial curvature fluctuations in such models is presented in the next section.

In single field inflation models where a potential term leads to the nearly constant vacuum energy and influences the spectrum of primordial fluctuations, the consistency relation links the spectral index of tensor perturbations and the tensor-to-scalar ratio. In the DHOST models that we will present here, the acceleration of the expansion of the Universe decouples from the perturbation generation. This implies that the consistency relation can be largely violated. We will find that the tensor-to-scalar ratio can be within reach of future experiments such as LITEBIRD even though the spectrum of tensor perturbation is almost perfectly scale-invariant. We also find that contrary to most single field models \citep{Martin:2024nlo}, the running of the spectral index is positive. 

Apart from the CMB, the early universe can be probed by using gravitational waves, which provide a complementary probe. Their recent discovery by LIGO \citep{LIGOScientific:2016aoc} has opened up a wealth of research avenues and can provide a unique link to higher energy and  particle physics \citep{Roshan:2024qnv}. Future gravitational wave detectors, both ground and space-based, will probe frequencies in the $\mu$Hz to kHz range (for interferometers) and up to the nanoHz range for pulsar timing arrays. These detectors can be used to constrain topological defects, such as cosmic strings \citep{LIGOScientific:2021nrg,Schmitz:2024hxw}, but they can also be employed to distinguish between various inflationary models. We will exploit this possibility and find that the spectrum of tensor perturbations, in the models that we consider, is so flat that detecting primordial gravitational waves coming from DHOST inflation in the mHz range may be conceivable in the distant future. This would allow one to disentangle models of inflation based on shift-invariant actions with a de Sitter background from single field models where the scalar potential is at the origin of both the cosmological background and perturbations.

In Section \ref{sec:model} we briefly describe the DHOST model, using the formalism developed in \citet{Brax:2021qlx}. In Section \ref{sec:gw} we present the formalism determining the gravitational wave density from inflationary models and compare the signatures of DHOST theories and slow-roll inflationary models. Finally we discuss the prospects of its detection with current and future gravitational wave detectors. We conclude in Section \ref{sec:concl}.

\section{Model}
\label{sec:model}

Considering only interactions up to second-order in the scalar field, the most general DHOST action can be written as
\begin{equation}
S = \int d^4 x \sqrt{-g} \Big[ F_0(\phi,X) + F_1(\phi,X) \Box \phi\nonumber     
\end{equation}
\begin{equation}
+ F_2(\phi,X) R + \sum_{i=1}^5 A_i(\phi,X) L_i \Big] \,,
\label{action-DHOST}    
\end{equation}
where $X=g^{\nu\eta}\phi_{\nu}\phi_{\eta}$, with $\phi_{\nu}\equiv\nabla_{\nu}\phi$. Here $L_i$ are  the five possible Lagrangians quadratic in the field $\phi$ and $A_i(\phi,X)$ their corresponding amplitudes with
\begin{align}\label{DHOST-L2s}
L_1 &= \phi_{\nu\eta} \phi^{\nu\eta} , \hspace{0.6cm} L_2 = (\Box \phi)^2 ,
\hspace{0.6cm} L_3 = \Box\phi \, \phi_{\nu}\phi^{\nu\eta} \phi_{\eta} , 
\nonumber \\
L_4 &= \phi^{\nu} \phi_{\nu\eta} \phi^{\eta\lambda}\phi_{\lambda} , 
\hspace{0.6cm} L_5 = (\phi_{\nu}\phi^{\nu\eta}\phi_{\eta})^2 \,.
\end{align}
By imposing no-ghost and gravitational waves constraints \citep{Crisostomi:2017pjs, Crisostomi:2019yfo} and simplifying the setting by assuming that the functions $F_i$ and $A_i$ only depend on the kinetic term $X$, the DHOST action becomes

\begin{align}
S_{\rm D} = \int d^4 x \sqrt{-g} \bigg[ F_0(X) + F_1(X) \Box\phi \nonumber \\
+ F_2(X) R
+ \frac{6 F_{2,X}^2}{F_2} \phi^{\nu} \phi_{\nu\eta} \phi^{\eta\lambda}\phi_{\lambda} \bigg]\,.
\label{action-dhost}
\end{align}
This action cannot be, however, used directly to build an inflationary model as the scalar power spectrum will always be scale-invariant \citep{Brax:2021qlx} and hence incompatible with the scalar spectral index measurements from inflation \citep{Akrami:2018odb}. This problem can be rectified by considering small shift-symmetry breaking perturbations of the function $F_0$ that can produce viable inflationary models \citep{Brax:2021qlx, Brax:2021bok, Sohn:2023dbp}.  

Here we consider a different scenario, by breaking the shift symmetry in the $F_2$ term. We add small perturbations of the function $F_2$ around the DHOST background action (\ref{action-dhost}), of the form
\begin{equation}
\label{F2pert}
F_2(X) \to F_2(X) -  \bigg[  \frac{m_{\rm phys}^2}{2}  \phi^2 + \frac{\lambda_{\rm phys}}{4!}  \phi^4 \bigg] \,.
\end{equation}
Here $F_2(X)$ is invariant under the rescaling $\phi \to \lambda \phi$, $x^\mu \to \lambda x^{\mu}$ which leaves $X$ invariant. 
The breaking of scale invariance is due to the two shift symmetry-breaking parameters $m_{\rm phys}$ and $\lambda_{\rm phys}$.

It is convenient to factor out the physical scales  and hence  re-express the action  in terms of tilded dimensionless coordinates \citep{Brax:2021qlx}
\begin{equation}\label{coordinets-ch}
{\tilde t} \equiv \Lambda t \,, \hspace{1cm} {\tilde x}^i \equiv \Lambda x^i \,,
\end{equation}
\begin{align}
\phi \equiv M \, \varphi\,, \hspace{.3cm} X\equiv{M^2 \Lambda^2}{\mathrm x}\,, \hspace{.3cm} \nonumber \\
F_0 \equiv \Lambda^4 f_0 \,, \hspace{.3cm}
F_1 \equiv \frac{\Lambda^2}{M} f_1 \,, \hspace{.3cm} F_2 \equiv \Lambda^2 f_2 \,,
\label{dimensionless-couplings}
\end{align}
and the dimensionless perturbation variables $m$ and $\lambda$,
\begin{equation}\label{dimensionless-couplings2}
m_{\rm phys} \equiv m \frac{\Lambda^2}{M}\,, \hspace{.5cm} \lambda_{\rm phys}\equiv\lambda\frac{\Lambda^4}{M^4} \,.
\end{equation}
The model depends on two mass scales, $M$ and $\Lambda$, where $\Lambda\simeq m_{\rm Pl}$. Here we consider the models as low energy effective theories well below the Planck scale, and hence it is not necessary to consider quantum gravity effects.

The background cosmology is determined by the shift-symmetric DHOST operators and leads to a de Sitter solution. The perturbation (\ref{F2pert}) to the $F_2$ term modifies the perturbation equations for the action (\ref{action-dhost}). These can be used to derive the scalar and tensor power spectra of curvature perturbations using the field quantisation procedure described in \citet{Gorji:2020bfl,Brax:2021qlx}. Compared to the previous work, these induce perturbations to two terms: the term multiplying the Ricci scalar and also the last terms of Eq. (\ref{F2pert}) through the denominator, as the term $F_{2,X}$ is unmodified. 

Such models can be parameterised at the perturbative level in terms of the coefficients  \citep{Bellini:2014fua, Gleyzes:2014rba, Langlois:2017mxy, Motohashi:2017gqb}
\begin{align}
&\alpha_H \equiv - {\rm x} \frac{f_{2,{\rm x}}}{f_2}\,, \hspace{1cm}
\alpha_B \equiv \frac{1}{2} \frac{\dot{\varphi}\,{\rm x}}{h_b} \frac{f_{1,{\rm x}}}{f_2} + \alpha_H \,, \hspace{1cm} \nonumber \\
&\alpha_K \equiv - \frac{{\rm x}}{6h_b^2}\frac{f_{0,{\rm x}}}{f_2} + \alpha_H + \alpha_B \,, \label{eqn:alpha_parameters}
\end{align}
depending on the first derivatives of the functions and on 
\begin{eqnarray}
&&\beta_K \equiv - \frac{{\rm x}^2}{3} \frac{f_{0,{\rm x}{\rm x}} }{h_b^2 f_2} 
+ (1-\alpha_H) (1+3 \alpha_B) + \beta_B \nonumber \\
&&\quad+ \frac{(1 + 6 \alpha_H - 3 \alpha_H^2) \alpha_K
- 2 ( 2 - 6 \alpha_H + 3 \alpha_K ) \beta_H}{1-3 \alpha_H} \,,
\nonumber \\
&& \beta_B \equiv  \dot{\varphi}\,{\rm x}^2 \frac{f_{1,{\rm x}{\rm x}}}{h_b f_2} \,,
\hspace{1cm}
\beta_H \equiv {\rm x}^2 \frac{f_{2,{\rm x}{\rm x}}}{ f_2}
\,,     \label{eqn:beta_parameters}
\end{eqnarray}
which are second order in derivatives.

Contrary to single-field inflationary models, the background and perturbative physics have different origins, i.e. they do not depend on the choice of a potential term for a normalised inflation field. Here,  inflation at the background level is due to solutions of the equations of motion with  a constant value of $X$. This leads to an inflationary phase. The deviations from a scale-invariant power spectrum at the perturbative level are due to the shift-symmetry parameters. 
Notice that  the mechanism leading to the end of inflation is not studied in this work. It could be related to a sudden change of physics at a particular time. However, as for the model described in Ref. \citet{Brax:2021qlx}, we can make an estimate of the number of e-foldings in the model between the time when the pivot scale $k_*$ enters the horizon and the end of inflation,
\begin{equation}
    N_*=\ln{\left(\frac{a_{\rm end}H_{\rm end}}{k_*}\right)} \,,
\end{equation}
obtaining a numerical result of $N_*=59.78$ for the parameter set described below.

The two Friedmann equations are obtained by considering the background Einstein equations of the action \eqref{action-dhost}. Then, in order to determine the scalar power spectrum, one needs to analyse the perturbations up to quadratic order of the action. After suitable integrations by parts and field redefinitions, the second-order action can be expressed in terms of a \textit{comoving curvature perturbation} $\zeta$ as
\begin{align}
\label{Lagrangian-DHOST-red}
{\tilde {\cal L}}_{\rm D}^{(2)}
= a^3 f_2 \bigg({\cal A} \, \dot{\zeta}^2 - {\cal B}\,  \zeta^2
\bigg) \,,
\end{align}
where ${\cal A}$ and ${\cal B}$ depend on the $\alpha_i$ and $\beta_i$ parameters and are given in appendix  \ref{sec:app}.

We seek inflationary solutions in a de Sitter phase of the early universe, where the Hubble parameter is constant. The Friedmann equation requires that
\begin{equation}\label{Stealth-Hubble}
h_{\rm dS} = \sqrt{\frac{-f_0}{6 f_2}} \,,
\end{equation}
and the scale factor is given in conformal time $\eta$ is
\begin{equation}
a(\eta)=-\frac{1}{h_{\rm dS} \eta}\,.   
\end{equation}
For simplicity, we are looking for linear solutions for the scalar field,
\begin{equation}
\varphi({\tilde t}) = c-{\tilde t}\, .
\end{equation}
For these solutions ${\mathrm x} = -1$ and therefore the functions $f_i$ and their derivatives are all evaluated on the background and become constants. Hence, the power spectra can be determined in terms of the $\alpha$ and $\beta$ parameters, which are also constants.  

{The power spectrum is obtained using the Mukhanov-Sasaki method for the quantisation of the fields, by considering the creation of primordial fluctuations from the Bunch-Davies vacuum. The second-order action can be re-expressed in conformal time as
\begin{equation}\label{Lagrangian-red-conf}
S^{(2)}_{\rm D}
= \int d\eta d^3  k  z^2  \,  \bigg[ \zeta'^2 -  {\cal C} \zeta^2
\bigg] \,,
\end{equation}
where ${\cal C}$ is a time-independent function of scale depending on $k$, ${\cal A}$ and ${\cal B}$.
To quantise the field, we then follow the procedure described in Ref. \citet{Peter:2013avv}. We define the modified Mukhanov-Sasaki variable as $v = z \zeta$, which satisfies a second-order differential equation, which can be solved with appropriate boundary conditions (a Bunch-Davies vacuum in the infinite past). Then the power spectrum is determined as
\begin{align}
\mathcal{P}_{\zeta}(k)&=\frac{k^3}{2 \pi^2} \left|\frac{v(k,\eta_*)}{z(k,\eta_*)} \right|^2,
\end{align}
where $\eta_*$ is the time corresponding to horizon crossing. Similarly, we determine the tensor power spectrum, with the coefficients corresponding to equation \eqref{Lagrangian-DHOST-red} given in the Appendix.} 
The role of the shift-symmetry breaking operators in $F_2$  is to modify the power spectrum of scalar and tensor perturbations, inducing small deviations from scale invariance.

The power spectrum of adiabatic scalar perturbations $\zeta$ has been measured with CMB probes, showing a small departure from scale invariance. This can be expressed as 
\begin{equation}
\mathcal{P}_{\zeta} (k) =  \mathcal{P}_{\zeta} (k_*)  \left(\frac{k}{k_*}\right)^{n_s-1} \, ,
\end{equation}
where $k_*$ is a pivot scale and $n_s$ is the spectral index. One can consider further variations of the spectral index as a series,
\begin{align}
\log \mathcal{P}_{\zeta} (k) =  \log \mathcal{P}_{\zeta} (k_*) 
+ \frac{1}{2} \frac{d \log n_s}{d \log k} \left(\frac{k}{k_*}\right)^2 \nonumber \\
+ \frac{1}{6} \frac{d^2 \log n_s}{d \log k^2} \left(\frac{k}{k_*}\right)^3 + \ldots \, .
\end{align}
These coefficients have been measured by \textit{Planck} with exquisite accuracy through the coefficients
\begin{align}
n_s({k}) & = 1 + \frac{d \log(\mathcal{P}_{\zeta}(k))}{d \log k}  \, , \\
\alpha_s (k) & = \frac{ d n_s(k)}{d \log k)} \, , \\
\beta_s (k) & = \frac{ d \alpha_s(k)}{d \log k} \, .
\end{align}
To build a viable inflationary model, we need to satisfy the latest inflationary constraints from the \textit{Planck} 2018 data release \citep{Akrami:2018odb} concerning the power spectrum: amplitude of the scalar power spectrum ($A_s$), scalar spectral index ($n_s$), running ($\alpha_s$) and running of the running of the spectral index ($\beta_s$), tensor-to-scalar ratio (defined as the ratio of the tensor power spectrum to the scalar one),

\begin{align}
\label{ns_const}
n_s &= 0.9625 \pm 0.0048    \, ,\\
\alpha_s &= 0.002 \pm 0.010 \, , \\
\beta_s & = 0.010 \pm 0.013 \, , \\
\ln(10^{10} A_s) & = 3.044 \pm 0.014 \,
\label{bs_const}
\end{align}
measured with $68\%$ confidence levels, using the TT,TE,EE+lowE+lensing likelihoods, at a pivot scale of $k_* = 0.05 \mathrm{~Mpc}^{-1}$. The observable scales today correspond to perturbations that have re-entered the Hubble radius around recombination, and hence $10^{-4}~\mathrm{Mpc}^{-1} \lesssim k \lesssim 10^{-1}~ \mathrm{Mpc}^{-1}$. In terms of the rescaled units of Eq. (\ref{coordinets-ch}), $k$ is such that
$\tilde{k}=\Lambda^{-1} k$.
The tightest constraints on the tensor-to-scalar ratio come form CMB experiments and are currently $r < 0.032$ at  95 \% confidence \citep{Tristram:2020wbi,Tristram:2021tvh} when combining \textit{Planck} with BICEP2/Keck 2015 data \citep{Ade:2018gkx}, further reduced to $r<0.028$ \citep{Galloni:2022mok} when adding LIGO-Virgo-KAGRA data \citep{KAGRA:2021kbb}.

\begin{figure*}
\includegraphics[width=0.9\linewidth]{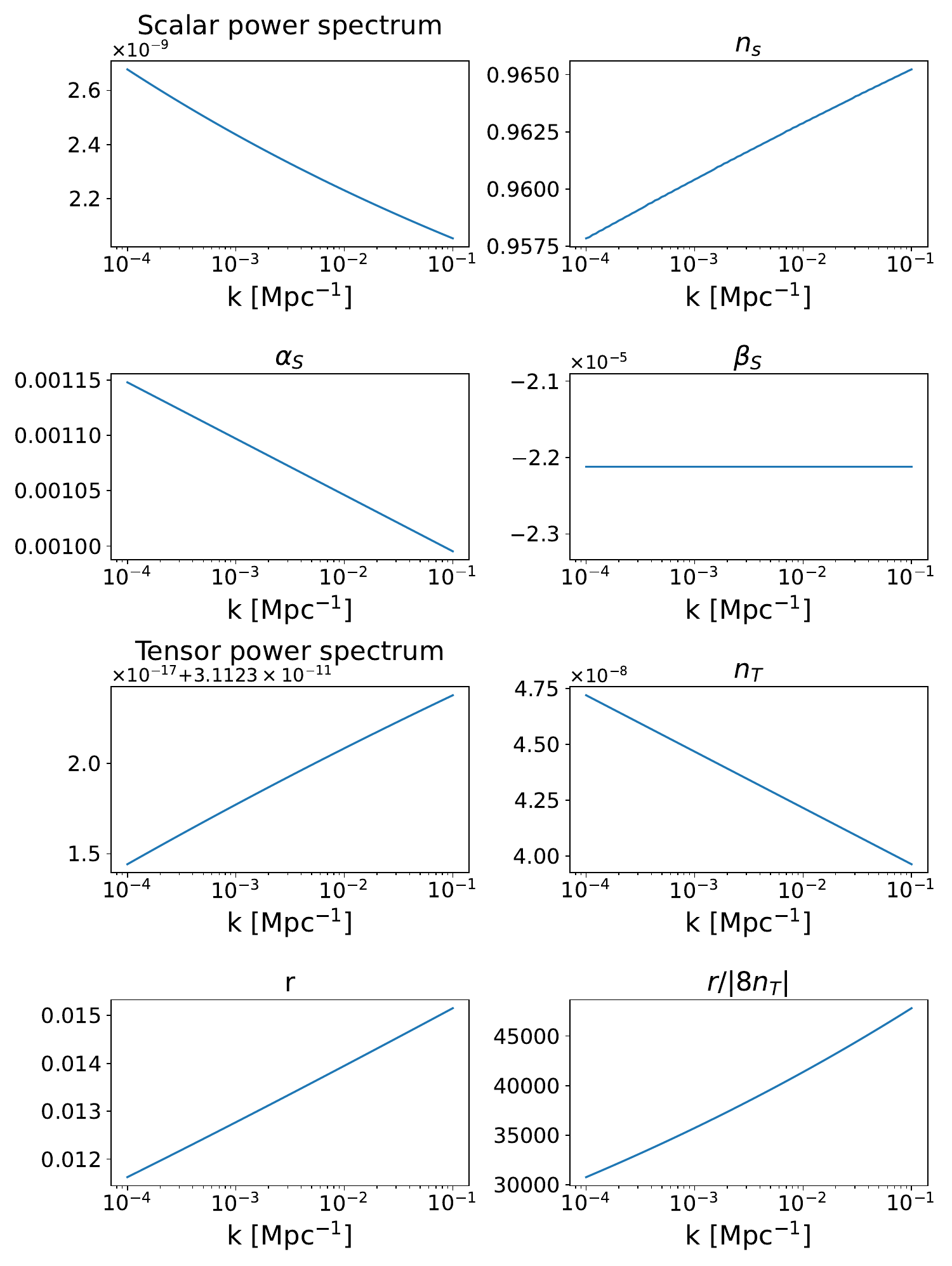} 
\caption{Model compatible with the \textit{Planck} 2018 data: from top to bottom and left to right, the scalar power spectrum, the scalar spectral index $n_s$, its running $\alpha_s$, the running of its running $\beta_s$, the tensor power spectrum, the tensor spectral index, the tensor-to-scalar ratio $r$,  and the value of $\frac{r}{8 |n_T|}$ for a DHOST model with $\alpha_H = 1.04$, $\alpha_B = 1$, $\beta_K = 3.97343$, $c = 1$, $h_{\rm dS} = 0.000050$, $f_2 = 4.34$ and the perturbation variables as $\lambda = -1.1 \times 10^{-30}$ and $m^2 = -1 \times 10^{-21}$. We can see that the consistency relation is maximally violated. }
\label{fig:res}
\end{figure*}

By choosing the model parameters $\alpha_H = 1.04$, $\alpha_B = 1$, $\beta_K = 3.97343$, $c = 1$, $h_{\rm dS} = 0.000050$, $f_2 = 4.34$ and the perturbation variables as $\lambda = -1.1 \times 10^{-30}$ and $m^2 = -1 \times 10^{-21}$, we obtain a model that is compatible with the above cited constraints.  We note that in our model the running of the spectral index is positive, which is different from the slow-roll inflationary models \citep{Martin:2024nlo}. This is another feature of DHOST inflation which differs from single field models where a potential is at the origin of both the de Sitter phase and the primordial fluctuations. The results are showed in Fig. \ref{fig:res}.  This choice is certainly not unique and a more thorough scan of the possibilities is left for future work.  Note the small values of the shift-symmetry coefficients whose tuning is not explained here and would need to be analysed in detail. This is left for future work too. 
We note that, while the scalar spectrum constraints are satisfied, and the tensor-to-scalar ratio $r$ is $r(k_*)=0.0147879$, smaller than the \textit{Planck} values, the tensor power spectrum varies only very little in the observable \textit{k}-range. We thus obtain a small tensor spectral index $n_T$ which induces a strong violation of the consistency relation $r\gg 8\vert n_T \vert$.

To get some insight into the parameter choices, we set $\alpha_B=1$. Other choices with $\alpha_B={\cal{O}}(1)$ could be chosen with similar results for the scalar to tensor ratio $r$ and the running of the spectral index. From the equations of the unperturbed model, we can derive $\alpha_H$ and $\beta_K$ by fixing  the sound speed of the scalar modes to be equal to the speed of light and by employing the constraint on $r$ as in \citep{Brax:2021qlx},
\begin{align}
\bar{c}_s^2&=\frac{(\alpha_H-\alpha_B)(1+\alpha_B)}{3 [(1+\alpha_B)^2-\beta_H]}\,, \\
r&=12 \bar{c}_s \frac{1-\frac{\beta_K}{(1+\alpha_B)^1}}{1+\frac{1}{\bar{c}_s^2}}\,.
\end{align}
The value of the Hubble parameter and of the $f_2$ parameter, together with the two perturbations, require tuning in order to get the correct amplitude and spectral index for the scalars, as they vary non-linearly.
To estimate the order of magnitude of the values of the perturbation parameters, we consider the Mukhanov-Sasaki equation for the normalised primordial fluctuations, which reads
\begin{equation}
v''(k,\eta)+K^2(k,\eta,\lambda,m^2)v(k,\eta)=0 \,,    
\end{equation}
where the primes represent derivatives with respect to conformal time. The linear term can be split into a background term, one depending on $\lambda$ and another on $m^2$, i.e. 
\begin{equation}
K^2(k,\eta,\lambda,m^2)=K^2_{\rm background}(k,\eta)+\lambda K^2_{\lambda}(k,\eta) +m^2 K^2_{m^2}(k,\eta)\,.    
\end{equation}
Due to the large values of the coefficients $K^2_{\lambda}(k,\eta)$ and $K^2_{m^2}(k,\eta)$, $\lambda \lesssim \mathcal{O}(10^{-30})$ and $m^2 \lesssim \mathcal{O}(10^{-18})$ ensure that the perturbative regime is indeed valid. In this range of parameters we find that the results are stable.

\section{Gravitational wave results}
\label{sec:gw}
This model, which violates the inflationary consistency relation, can provide a distinct phenomenology compared to the standard slow-roll single-field inflationary model in the frequency range that can be probed by gravitational wave detectors. In this section, we describe the gravitational wave predictions of the model.

The gravitational wave density parameter can be determined as \citep{Clarke:2020bil},

\begin{equation} \label{eqn:omgw_convert}
\Omega_{\rm GW} (k)h^2 = \frac{3}{128} \Omega_\text{r} h^2 \mathcal{P}_T(k) \left[ \frac{1}{2} \left(\frac{k_\text{eq}}{k}\right)^2 +\frac{16}{9}\right] \,,
\end{equation}
where $\Omega_{\rm r}$ is the radiation density, $\mathcal{P}_T(k)$ is the tensor power spectrum and $k_\text{eq}$ is the wavenumber corresponding to radiation-matter equality. This can be expressed in terms of the matter density $\Omega_\text{m}$ and the speed of light $c_l$ as
$k_\text{eq}=\sqrt{2}H_0\Omega_\text{m}/(c_l\sqrt{\Omega_\text{r}})$. In the case of a single-field slow-roll model, where the consistency relation is satisfied ($n_T=-r/8$), the tensor power spectrum becomes
\begin{equation} \label{eq:tensorpower}
    \mathcal{P}_T(k)=r A_{\rm s} \left(\frac{k}{k_*}\right)^{n_T} \,.
\end{equation}
The conversion between scales and frequencies can be obtained as
\begin{equation}f=(1.55\times10^{-15} \, \text{Hz Mpc})\times k \,.
\end{equation}
We consider the density of gravitational waves arising from the model presented in the previous section and the one from a single-field slow-roll inflationary model with the same tensor-to-scalar ratio $r$. 

The scale dependence of the gravitational wave density parameter of the DHOST model and the slow-roll one are qualitatively different: in the case of the DHOST theories, it is constant, while for the slow-roll model it is slowly decaying and has a lower overall amplitude in the frequency range of the gravitational wave detectors. 

\begin{figure}
\includegraphics[width=0.99\linewidth]{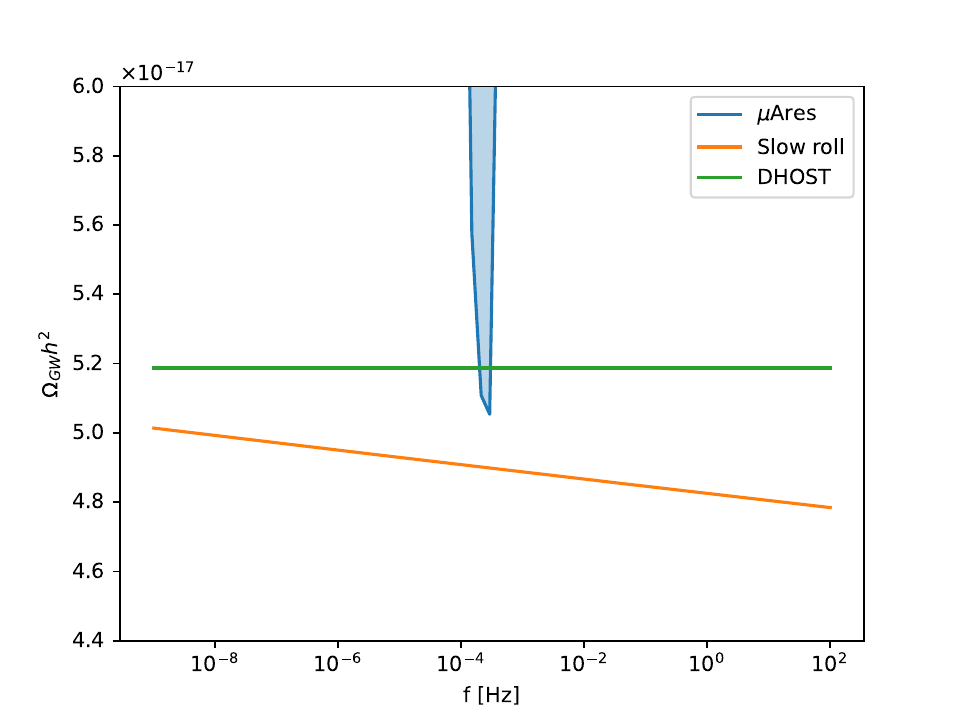} 
\caption{Gravitational wave density parameter for the slow-roll and DHOST model together with the power law integrated sensitivity curves for $\mu$ARES from Ref. \citet{Roshan:2024qnv}.}
\label{fig:gw}
\end{figure}

The DHOST model predicts a constant $\Omega_{\rm GW}h^2 = 5.18 \times 10^{-17}$, which is currently below the sensitivity bounds of current gravitational wave experiments \citep{Roshan:2024qnv}. However, it is very close to the sensitivity threshold of the proposed space-based interferometer $\mu$ARES \citep{Sesana:2019vho}, which will be able to distinguish between the DHOST and slow-roll model in the mHz frequency range, {as illustrated in Fig. \ref{fig:gw}. The orange curve represents a slow-roll single field model which has the same tensor-to-scalar ratio as the DHOST one, and satisfies $n_T=-r/8$.}

\section{Conclusions}
\label{sec:concl}
In this work we described how second-order scale-invariant DHOST theories  perturbed by shift-symmetry breaking terms multiplying the Ricci scalar whilst allowing for inflationary solutions in the early universe in a de Sitter phase can lead to a nearly flat gravitational wave spectrum. For the perturbed models, the scalar and tensor power spectra can be  explicitly shown to be  compatible with the latest constraints from \textit{Planck} at the power spectrum level  and  provide a tensor-to-scalar ratio compatible with observations. We find that the running of the scalar spectral index is positive, a significant difference with the expected result for single field inflation \citep{Martin:2024nlo}.  These perturbed model also provide a strong violation of the consistency relation for slow-roll single-field inflationary models and  a distinct signature in the gravitational wave sector.  These are the two striking differences between our scenario and the standard lore.   By computing the gravitational wave density parameter for the DHOST model and for a slow-roll one having the same tensor-to-scalar ratio, we showed that the DHOST model has a constant density, while for the slow-roll model this has a lower amplitude and is slowly decaying. The amplitude of the DHOST model is within the detectable range of the future $\mu$ARES space-based interferometer. Such probes should thus be able to distinguish between single-field inflationary scenarios where the inflaton potential generates both the initial de Sitter phase and the breaking of scale invariance for the scalar and tensor spectra,  and DHOST models with shift-symmetry breaking operators where the tensor power spectrum is nearly constant. We note that the results are based on a subset of DHOST theories, which have a background depending on the kinetic term $X$,  with small perturbations breaking  scale invariance and a linear solution in time for the scalar field. Different assumptions in the model set-up could yield different phenomenologies. This is left for subsequent work.

\section*{Acknowledgements}

We would like to thank C. Ringeval for crucial discussions and M. Tristram for relevant remarks. AL is supported by a United Kingdom Research and Innovation (UKRI) Future Leaders Fellowship [Grant No. MR/V021974/2].

\section*{Data Availability}
No data were created or analysed in this study.



\bibliographystyle{mnras}
\bibliography{Bibliography} 




\appendix

\section{Functions of the second order expansion}
\label{sec:app}

The functions ${\cal A}$ and ${\cal B}$ are given in a de Sitter scenario for scalar and tensor perturbations as
${\cal A}={\cal A}_0+{\cal A}_{\rm pert}$ and ${\cal B}={\cal B}_0+{\cal B}_{\rm pert}$, with

\begin{align}
{\cal A}_0^{\rm scalar}&=    6 \bigg[ 1 - \frac{\beta_K}{(1+\alpha_B)^2} \bigg] \,, \\
{\cal B}_0^{\rm scalar}&=  \frac{2 k^2}{a^2} 
\frac{\alpha_B - \alpha_H }{1+\alpha_B}  \,,   
\end{align}

\onecolumn
\begin{align}
{\cal A}&_{\rm pert}^{\rm scalar}=\frac{\phi ^2 \left(\lambda  \phi ^2+12 m^2\right)}{3072  (\alpha_B+1)^4 f_2}  
\times \Bigg(-\frac{128 (\alpha_B+1)^2 \left(a^2 \left(54 \alpha_H
^2 (f_2-1) h_{\rm dS}^2+9 \left(\alpha_H^2-1\right) h_{\rm dS}^2\right)+2 \alpha_H k^2 (\alpha_H-3 \alpha_H f_2-2)\right)}{h_{\rm dS}^2} \nonumber \\
&+\frac{1}{3 a^2 \beta_K h_{\rm dS}^3 \phi ^2 \left(\lambda  \phi
   ^2+12 m^2\right)-16 (\alpha_B+1)^2 f_2 h_{\rm dS} k^2} \bigg(768 a^2 (\alpha_B+1)^3 \left(3 a^2 \beta_K
   h_{\rm dS}^2 \phi ^2 \right. \nonumber \\
&\times \left.(6 \alpha_H h_{\rm dS}+h_{\rm dS} (\alpha_H (9 h_{\rm dS}-3)-2)) \left(\lambda  \phi
   ^2+12 m^2\right)+16 f_2 k^2 \left(2 \beta_K (h_{\rm dS}-3 \alpha_H h_{\rm dS}) \right.\right.\nonumber \\
   &\left.\left.-3 (\alpha_B+1)^2
   \alpha_H h_{\rm dS} (3 h_{\rm dS}-1)\right)\right)\bigg) \nonumber \\
   &+\frac{1}{{\left(f_2 k^2-\frac{3 a^2 \beta_K h_{\rm dS}^2
   \left(\lambda  \phi ^4+12 m^2 \phi ^2\right)}{16 (\alpha_B+1)^2}\right)^2}} 
   \bigg(3 a^2 \left(9 a^4 \beta_K^2 h_{\rm dS}^4 \phi
   ^4 \left(\lambda  \phi ^2+12 m^2\right)^2-96 a^2 f_2 h_{\rm dS}^2 k^2 \right.\nonumber \\
   &\times \left.\left(12 f_2 \left((\alpha_B+1)^2-\beta_K\right)^2+\beta_K^2 \phi ^2 \left(\lambda  \phi ^2+12 m^2\right)\right)-256 (\alpha_B+1)^2 f_2^2 k^4
   \left((\alpha_B+1)^2-2 \beta_K\right)\right)\bigg)\Bigg)  \,, 
\end{align}

\begin{align}
{\cal B}_{\rm pert}^{\rm scalar}&=-\frac{\phi }{{72 a^4 (\alpha_B+1)^2 f_2 h_{\rm dS}^2}} 
\times  \Bigg[\frac{1}{\left(\alpha_B^2+2 \alpha_B-\beta_K+1\right) \left(16 (\alpha_B+1)^2 f_2 k^2-3 a^2
   \beta_K h_{\rm dS}^2 \phi ^2 \left(\lambda  \phi ^2+12
   m^2\right)\right)^2} \nonumber\\
   &\times \left(768 a^2 (\alpha_B+1)^3 f_2^2 h_{\rm dS} k^4 \left(9 a^2
   h_{\rm dS} \left(\alpha_B^2+2 \alpha_B-\beta_K+1\right)
   \left(h_{\rm dS} \left(\alpha_B^2 (3 \alpha_H+2) + \alpha_B (6 \alpha_H+4)\right. \right.\right.\right. \nonumber\\
   &-\left.\left. \alpha_H \beta_K+3 \alpha_H-\beta_K+2\right)-(\alpha_B+1)^2 \alpha_H\right)
   \left(\lambda  \phi ^2 (3 h_{\rm dS} \phi -4) +12 m^2 (3 h_{\rm dS} \phi -2)\right) \nonumber \\
   &+2
   (\alpha_B+1) k^2 \left(2 \alpha_B^3+\alpha_H^2 (3-4
   \beta_K)-2 \alpha_H (\beta_K-1)-\beta_K \right. +\alpha_B^2 \left(2 \alpha_H-3 \alpha_H^2
   (f-2)+5\right)\nonumber \\
   &+\left.\alpha_B \left(\alpha_H (\beta_K+4)-2
   \beta_K+3 \alpha_H^3 (f-1) \right.
   \left.\left.\left.-3 \alpha_H^2 (f-3)+4\right)+3
   \alpha_H^3 (f-1)+1\right) \left(\lambda  \phi ^2 (h_{\rm dS} \phi -4)+12 m^2
   (h_{\rm dS} \phi -2)\right)\right)\right) \nonumber \\
   &+\phi  \left(\lambda  \phi ^2+12 m^2\right) \Big(3 a^2
   h_{\rm dS}^2 \left(-\left((\alpha_B+1)^2 \left(9 a^2 h_{\rm dS}^2+2
   k^2\right)\right)-3 (\alpha_H+1)^2 k^2\right) \nonumber\\
   &-\frac{2 (\alpha_B+1)^2
   \alpha_H^2 (f_2-1) k^4 (\alpha_B-\alpha_H)^2}{\left(\alpha_B^2+2 \alpha_B-\beta_K+1\right)^2}+\frac{4 (\alpha_B+1) \alpha_H (\alpha_H+1)
   k^4 (\alpha_B-\alpha_H)}{\alpha_B^2+2 \alpha_B-\beta_K+1}\Big)\Bigg] \,,
\end{align}
and
\begin{align}
{\cal A}_0^{\rm tensor}&=    1 \,, \\
{\cal B}_0^{\rm tensor}&= \frac{k^2}{a^2}  \,,  \\
{\cal A}_{\rm pert}^{\rm tensor}&= -\frac{\phi^2 (12 m^2+\lambda \phi^2)}{24 f_2}  \,, \\
{\cal B}_{\rm pert}^{\rm tensor}&= -\frac{\phi ^2 \left(6 a^2 h_{\rm dS}^2+k^2\right) \left(\lambda \phi ^2+12
   m^2\right)}{24 a^2 f_2}  \,.   
\end{align}


\bsp	
\label{lastpage}
\end{document}